\documentclass[showpacs,amsmath,amssymb, nobibnotes, aps, prl,showkeys, floatfix]{revtex4}
\usepackage{graphicx}

\usepackage{graphicx}
\begin{document}

\title{ Influences of Dark Energy and dark matter on Gravitational Time Advancement}
\author{ Samrat Ghosh$^1$ and Arunava Bhadra$^1$}
 \affiliation{ $^{1}$ High Energy $\&$ Cosmic Ray Research Center, University of North Bengal, Post N.B.U, Siliguri 734013, India. }

\begin{abstract}
The effect of dark matter/energy on gravitational time advancement (negative effective time delay) has been investigated considering few dark energy/matter models including cosmological constant. It is found that dark energy gives only (positive) gravitational time delay irrespective of the position of the observer whereas pure Schwarzschild geometry leads to gravitational time advancement when the observer is situated at relatively stronger gravitational field point in the light trajectory. Consequently, there will be no time advancement effect at all at radial distances where gravitational field due to dark energy is stronger than the gravitational field of Schwarzschild geometry.    
\end{abstract}
\date{24-08-2015}
\pacs{98.62.Mw, 04.70.Bw, 95.30.Sf, 04.20.-q, 04.20.DW}
\keywords{relativity and gravitation, classical general relativity}
\maketitle

\section{Introduction}
The discovery of the acceleration of the universe's expansion [1-5] has led inclusion of a new component into the energy-momentum tensor of the universe having negative pressure, the so called dark energy component. On the other hand data from rotation curve surveys [6] and few other observations [7,8] demand that there is a dominating component of matter in galaxies which is non-luminous or dark. Several other observations which include the cosmic microwave background (CMB) measurements [9-12], baryon acoustic oscillations (BAO) [13-15], lensing in clusters [16,17] support the existence of dark energy as well as the presence of dark matter halo surrounding the Galactic disc. Consequently on large distance scales, astrophysical and cosmological phenomena are governed mainly by dark matter and dark energy. 

The simplest candidate for dark energy is the cosmological constant ($\Lambda$): a constant energy density with equation-of-state parameter w=-1 and the $\Lambda$CDM model where CDM refers to cold dark matter, is in accordance with all the existing cosmological observations [18,19] such as the cosmic microwave background (CMB) anisotropies, the large scale structure, the scale of the baryonic acoustic oscillation in the matter power spectrum and the luminosity distance of the supernovae type Ia but it has a big theoretical problem - it's size ($\sim 10^{-52} \; m^{-2}$) is many orders of magnitudes below the expected vacuum energy density in the standard model of particle physics [20]. Hence many other theoretical explanations for the DE have been proposed in the literature in which the parameter w evolves with time or different from $-1$ such as the quintessence [21-23], k-essence [24-27], phantom field [28,29], Chaplygin gas [30,31] models. There are also proposals for modification of general relativity which include Scalar tensor theories [32] or f(R) gravity models [33], conformal gravity model [34,35], massive gravity theories [36] including Dvali-Gabadadze-Porrati (DGP) braneworld gravity [37,38] models etc., which lead to late-time accelerated expansion without invoking any dark energy. 

Like dark energy, there are also several candidates for dark matter [39] such as WIMPs, Axions, Sterile neutrinos etc. There are proposals for modifications at the fundamental theoretical level as well which include MOND [40-43] that suggests modifications in Newtonian dynamics. The evidence of presence of non-baryonic dark matter from the CMB data, however, questions the MOND like schemes. The conformal gravitational theory [34, 35], which is based on Weyl symmetry, also can explain flat rotation curves of galaxies without the need of dark matter.  

Dark energy/matter is likely to affect the gravitational phenomena in all distance scales including the local scales. Several investigations have so far been made to estimate the influence of dark energy (mainly through cosmological constant) on different local gravitational phenomena which include the three classical observables - the perihelion shift of planets [44, 45], gravitational bending of light [45-49] and gravitational time delay (or Shapiro time delay) [45, 50.51]. Due to the tiny value of $\Lambda$, the influence of dark energy has been found very small, not detectable by the ongoing experiments. Out of the local gravitational phenomena the effect of $\Lambda$ is found maximum in the case of perihelion precession of planets and the observations on perihelion precession of mercury put an upper bound of $\Lambda \le 10^{-42} \; m^{-2}$ [51]. On the other hand analysis of the perihelion precession of Mercury, Earth, and Mars also lead to a upper bound $ 3 \times 10^{-19}$ $g/cm^{3}$ for dark matter density ($\rho_{dm}$) [53] whereas the rotation curve data implies that $\rho_{dm}$ in the Milky Way at the location of solar system is $\rho_{dm} = 0.5 \times 10^{-24}$ $g/cm^{3}$ [54]. 

In this work we would like to examine the influence of dark energy and dark matter on gravitational time advancement. Gravitational time advancement effect takes place when the observer is situated at stronger gravitational field in respect to the gravitational field encountered by the photon while traversing to a certain path [55]. We found that dark energy and dark matter do affect the gravitational time advancement and though the magnitude of the effect is small, it induces an interesting observational consequence, at least in principle.

The organization of the paper is the following. In the next section we discuss briefly the gravitational time advancement effect. The influence of Dark energy and dark matter on Gravitational Time Advancement are evalued in Section 3. The results are discussed in section 4 and finally we conclude in section 5.           

\section{Gravitational Time Advancement}
The gravitational time delay is one of the classical solar system tests of General relativity. The general perception about gravitational time delay is that due to influence of gravitating object the average global speed of light is reduced from its spacial relativistic value $c_0$ and hence the signal always suffers an additional time delay. But depending upon the position of the observer, the delay can as well be negative which was mentioned as gravitational time advancement [55]. To exemplify the effect let us consider light is propagating in a gravitational field between two points A and B. Assuming standard Schwarzschild geometry, i.e. 

\begin{equation}
ds^2=-(1-2\mu/r) dt^2+(1-2\mu/r)dr^2+r^2/d{\Omega}^2 \; ,
\end{equation} 

total coordinate time required for the round trip journey between the points A and B (or between the points B to A and back) to the first order in $\mu=GM/c_0^2$, is given by [14]

\begin{eqnarray}
c_0 \Delta t_{AB}&= 2\left(\sqrt{r_A^2-r_o^2}+\sqrt{r_B^2-r_o^2}\right) 
+4\mu \left(ln\frac{r_A+\sqrt{r_A^2-r_o^2}}{r_o} + ln\frac{r_B+\sqrt{r_B^2-r_o^2}}{r_o} \right) + 
2\mu \left[\left(\frac{r_A-r_o}{r_A+r_o}\right)^{1/2} +\left(\frac{r_B-r_o}{r_B+r_o}\right)^{1/2} \right]
\end{eqnarray}

where $r_A$ and $r_B$ are the radial coordinates of the point A and B respectively and $r_o$ is the closest distance to the gravitating object in the photon path.  

Suppose the point A is located at relatively much weaker gravitational field due to mass M than the point B i.e. $r_A >> r_B$ where $r_A$ and $r_B$ are the values of coordinate r evaluated at the position A and B respectively. Hence the proper time for between transmission and the reception of the signal to be measured by the observer at the point A is 

\begin{eqnarray}
c_0\Delta\tau_{AB}\simeq(1-\frac{\mu}{r_A})\Delta t_{AB}\simeq 2\left(\sqrt{r_A^2-r_o^2}+\sqrt{r_B^2-r_o^2}\right) 
+4\mu \left(ln\frac{r_A+\sqrt{r_A^2-r_o^2}}{r_o} + ln\frac{r_B+\sqrt{r_B^2-r_o^2}}{r_o} \right) +\nonumber \\
2\mu \left[\left(\frac{r_A-r_o}{r_A+r_o}\right)^{1/2} +\left(\frac{r_B-r_o}{r_B+r_o}\right)^{1/2} \right] \;.
\end{eqnarray}

In the above expression the first term on the right hand side is the usual the special-relativistic time of travel. The rest two terms are general relativistic corrections. As a result the observed time will be higher than the time taken between transmission and the reception in the absence of gravitating object which is the well known gravitational time delay. 

Now let us consider the case that the observer is at the point B instead of the point A. In that case the proper time between transmission and the reception of the signal to be measured by the observer will be [55]

\begin{eqnarray}
c_0\Delta\tau_{AB}\simeq(1-\frac{\mu}{r_B})\Delta t_{AB} \simeq 2\left(\sqrt{r_A^2-r_o^2}+\sqrt{r_B^2-r_o^2}\right) 
+4\mu \left(ln\frac{r_A+\sqrt(r_A^2-r_o^2)}{r_o} + ln\frac{r_B+\sqrt(r_B^2-r_o^2)}{r_o} \right) + \nonumber \\
2\mu \left[(\frac{r_A-r_o}{r_A+r_o})^{1/2} +(\frac{r_B-r_o}{r_B+r_o})^{1/2} \right]
-2\mu(\frac{\left(\sqrt{r_A^2-r_o^2}+\sqrt{r_B^2-r_o^2}\right)}{r_B})&
\end{eqnarray}

Due to the last term of the right hand side of the above expression, which is the dominating one among the general relativistic correction terms, the time taken between transmission and the reception will be reduced from the usual the special-relativistic time of travel when the distance between A and B exceeds certain value. This effect is known as gravitational time advancement (negative time delay) that arises because of the clock runs differently at different positions in gravitational field.  

\section{Influence of Dark energy on Gravitational Time Advancement}
In the presence of dark energy the exterior spacetime of spherically symmetric mass distribution is no longer described by Schwarzschild geometry, but by some modification of Schwarzschild metric. For instance if dark energy is cosmological constant, the exterior static spacetime will be Schwarzschild- de Sitter (SDS) spacetime. 

Here we shall consider a general static spherically symmetric metric of the form

\begin{eqnarray}
ds^2=-B(r)dt^2+A(r)dr^2+r^2/d{\Omega}^2&
\end{eqnarray}

with
\begin{equation}
B(r)=1-2m/r-a\Lambda r^{n}/3
\end{equation}
and
\begin{equation}
A(r)=(1-2m/r-\Lambda r^{n}/3)^{-1}
\end{equation}

where $a$ and $\Lambda$ are constants. Different choices of n and a lead to different models of dark energy. 

case 1: With n=1/2, a=2, $\Lambda=\pm \sqrt{GM/r_{c}^{2}}$ the model represents the gravitational field of a spherically symmetric matter distribution on the background of an accelerating universe in Dvali-Gabadadze-Porrati (DGP) braneworld gravity provided leading terms are only considered [56]. $r_{c}$ is the crossover scale beyond which gravity becomes five dimensional.  

case 2: For the choice n=1, a=1 and negative $\Lambda$, the model well describes the gravitational potential due to central matter distribution plus dark matter [34, 35, 57]. 

case 3: If n=3/2, a=2/3 and $\Lambda=-m_{g}^{2}\sqrt{\frac{2GM}{13c^{2}}}$, the model corresponds to the non-perturbative solution of a massive gravity theory (an alternative description of accelerating expansion of the universe) [58] where $m_{g}$ is the mass of graviton. 

case 4: When a=1, n=2 and $m=\mu$ the above metric describes the Schwarzschild-de Sitter (SDS) or Kotler space-time which is the exterior space time due to a static spherically symmetric mass distribution in presence of the cosmological constant $\Lambda$ [59].

\subsection{General trajectory}
Now let us suppose that a light beam is moving between two points A and B in the gravitational field of Eqs.(5-7). The expression for coordinate time required for light rays to traverse the distance $r_{o}$ to $r$, where $r_{o}$ is the closest distance from the gravitating object over the trajectory can be obtained from geodesic equations which is given by 
 
\begin{equation}
\delta t=\int_{r_o}^{r}\sqrt{P(r,r_{o})} \, dr \, , 
\end{equation}
where, 

\begin{equation}
P(r,r_{o})=\frac{A(r)/B(r)}{1-\frac{r_{o}^{2}}{r^{2}}\frac{B(r)}{B(r_{o}}}
\end{equation}

For general power index ($n$) of $\Lambda$ in Eq.(5), the above equation after integration can only be expressed in terms of hyper-geometric functions and thereby not very useful. However, for n=1 and n=2, the integral can be written in a handful form, particularly when higher order terms in M and $\Lambda$ are ignored. The extra coordinate time delay ($\delta t^{\Lambda}_{1}$) induced by the dark sector terms in Eq.(8) is given by for n=1 and $\Lambda=-\Lambda$,
  
\begin{equation}
\delta t^{\Lambda}_{1}=-(a+1)\frac{\Lambda}{12} \left(r\sqrt{r^2-r_{o}^{2}} +r_{o}^2 ln(r+\sqrt{r^2-r_0^2})\right) - \frac{a\Lambda r_{o}^{2}}{6} \left(ln(r+\sqrt{r^2-r_0^2})-\sqrt{\frac{r-r_{o}}{r+r_{o}}}\right)  \; , 
\end{equation}    

while for n=2
\begin{equation}
\delta t^{\Lambda}_{2}=(a+1)\frac{\Lambda}{18} \left((r^2+2r_{o}^{2})\sqrt{r^2-r_{o}^{2}} \right) - \frac{a\Lambda r_{o}^{2}}{6}\sqrt{r^2-r_{o}^{2}}   \; ,
\end{equation}    

and for general n ($n \ne 1$) when $r_A >> r_o$ and $r_B >>r_o$, 

\begin{eqnarray}
\delta t^{\Lambda}_{n} \simeq  \frac{(a+1) \Lambda}{6(n+1)} r^{n+1} - \frac{(a-1)\Lambda }{12(n-1)} r^{n-1} r_{o}^{2} + O (r_{o}^{4})
\end{eqnarray}

Hence the proper time between the transmission and the reception of the signal to be measured by the observer at point B will be for n=1

\begin{eqnarray}
c_0 \Delta\tau_{1} &\simeq& 2\left(\sqrt{r_A^2-r_o^2}+\sqrt{r_B^2-r_o^2}\right)  +4\mu \left(ln\frac{r_A+\sqrt{r_A^2-r_o^2}}{r_o} + ln\frac{r_B+\sqrt{r_B^2-r_o^2}}{r_o} \right) + 2\mu \left[(\frac{r_A-r_o}{r_A+r_o})^{1/2} +(\frac{r_B-r_o}{r_B+r_o})^{1/2} \right]  \nonumber \\ 
&& -(a+1)\frac{\Lambda}{12} \left(r_{A}\sqrt{r_{A}^{2}-r_{o}^{2}} +r_{o}^2 \; ln(r_{A}+\sqrt{r_{A}^2-r_{o}^{2}}) \right) + \frac{a\Lambda r_{o}^{2}}{6} \left(\sqrt{\frac{r_{A}-r_{o}}{r_{A}+r_{o}}}-ln(r_{A}+\sqrt{r_{A}^2-r_{o}^{2}})\right) - \nonumber \\
&& (a+1)\frac{\Lambda}{12} \left(r_{B}\sqrt{r_{B}^{2}-r_{o}^{2}} +r_{o} ln(r_{B}+\sqrt{r_{B}^2-r_{o}^{2}}) \right) + \frac{a\Lambda r_{o}^{2}}{6} \left(\sqrt{\frac{r_{B}-r_{o}}{r_{B}+r_{o}}}-ln(r_{B}+\sqrt{r_{B}^2-r_{o}^{2}})\right) \nonumber \\
&&-2\left(\frac{\mu}{r_B}-\frac{a\Lambda r_{B}}{3}\right)\left(\sqrt{r_A^2-r_o^2}+\sqrt{r_B^2-r_o^2}\right)
\end{eqnarray}

Usually for observing time advancement effect, $r_{o}=r_{B}$. Further for describing flat rotation curve, a has been chosen as $1$. Hence the above equation reduces to
\begin{eqnarray}
c_0 \Delta\tau_{1} &\simeq& 2\sqrt{r_A^2-r_B^2} + 4\mu \;ln\left(\frac{r_A+\sqrt{r_A^2-r_B^2}}{r_B} \right) +  2\mu \left(\frac{r_A-r_B}{r_A+r_B}\right)^{1/2} \nonumber \\
&& - \frac{\Lambda}{6} \left[\left(r_{A}\sqrt{r_{A}^{2}-r_{B}^{2}} +r_{B}^2 \;ln \left(r_{A}+\sqrt{r_{A}^2-r_{B}^{2}}\right) \right) - r_{B}^{2} \left(\sqrt{\frac{r_{A}-r_{B}}{r_{A}+r_{B}}}-ln(r_{A}+\sqrt{r_{A}^2-r_{B}^{2}})\right) \right] \nonumber \\
&& -2\left(\frac{\mu}{r_B}-\frac{\Lambda r_{B}}{3}\right)\sqrt{r_A^2-r_B^2}
\end{eqnarray}

When $r_{A} >>r_{B}$, the above equation transforms to
\begin{eqnarray}
c_0 \Delta\tau_{1} &\simeq& 2 r_A - 2 \mu \left(\frac{r_{A}}{r_{B}} - 2ln\left(\frac{2r_A}{r_B} \right) -1 \right) - \frac{\Lambda}{6}\left(r_{A}^{2} + 2 r_B^2 ln2r_A  
-4 r_{A} r_{B}  \right)
\end{eqnarray}

Similarly for n=2 with $r_{o}=r_{B}$ 
\begin{eqnarray}
c_0 \Delta\tau_{2} &\simeq& 2\sqrt{r_A^2-r_B^2} + 4\mu \;ln\left(\frac{r_A+\sqrt{r_A^2-r_B^2}}{r_B} \right) +  2\mu \left(\frac{r_A-r_B}{r_A+r_B}\right)^{1/2} + (a+1)\frac{\Lambda}{18} \left(r_{A}^{2}+2r_{B}^{2}\right)\sqrt{r_{A}^{2}-r_{B}^{2}} - \nonumber \\
&& \frac{a\Lambda r_{B}^{2}}{6} \sqrt{r_{A}^{2}-r_{B}^{2}} -2\left(\frac{\mu}{r_B}+\frac{a\Lambda r_{B}^{2}}{3}\right)\sqrt{r_A^2-r_B^2}
\end{eqnarray}

which for $r_{A} >>r_{B}$ becomes

\begin{eqnarray}
c_0 \Delta\tau_{2} &\simeq& 2 r_A + 2 \mu \left(2ln\left(\frac{2r_A}{r_B} \right) + 1 -\frac{r_{A}}{r_{B}}\right) + \frac{\Lambda}{18}\left((a+1)r_{A}^{3} + a r_{A} r_{B}^{2} \left(2-13a)\right)  
-2\frac{a r_{A} r_{B} }{3} \right)
\end{eqnarray}

and for general n
\begin{eqnarray}
c_0 \Delta\tau_{n} &\simeq& 2 r_A + 2 \mu \left(2ln\left(\frac{2r_A}{r_B} \right) + 1 -\frac{r_{A}}{r_{B}}\right) + \frac{(a+1) \Lambda}{6(n+1)} r^{n+1} - \frac{(a-1)\Lambda }{12(n-1)} r^{n-1} r_{o}^{2} - \frac{2a\Lambda r_{A} r_{B}^{n}}{3}  
\end{eqnarray}

Unless $\Lambda$ effect dominates over the pure Schwarzschild effect, the net time delay will be negative in all the above cases resulting time advancement. 

\subsection{Small distance travel}
Let us suppose a light beam is moving from a point on the Earth surface(B) $(R_E,\theta,\phi)$ where the radius of Earth is denoted as $R_E$, to a nearby point with corrdinates C$(R_E+\Delta R_{E},\theta,\phi)$ and reflects back to the transmitter position(B). The light signal will travel null curve of space-time satisfying  $ds^2=0$. Then the proper distance between point B and point C is given by,

\begin{eqnarray}
&\Delta L_{BC}=\int_{R_1}^{R+\Delta{R}}(1-2m/r-{\Lambda{r}^n}/3)^{-1/2}dr \nonumber \\
               &\simeq\Delta{R}[1+\frac{m}{R}-\frac{m\Delta{R}}{2R^2}+\frac{\Lambda{R}^n}{6}(1+\frac{n\Delta{R}}{2R}+\frac{n(n-1)\Delta{R}^2}{6R^2}) \nonumber \\&
+\frac{3m^2}{2R^2}+\frac{m\Lambda{R}^{n-1}}{2}(1+\frac{(n-1)\Delta{R}}{2R})]&
\end{eqnarray}

The coordinate time interval in transmitting a light signal from B to C and back, is given by,

\begin{eqnarray}
&{\Delta{t}}=2\int_{R_1}^{R+\Delta{R}}{(1-\frac{2m}{r}-\frac{\Lambda{r}^n}{3})^{-1/2}(1-\frac{2m}{r}-\frac{a\Lambda{r}^n}{3})^{-1/2}}dr \nonumber\\
           &\simeq2L_{BC}[1+\frac{m}{R}+\frac{3m^2}{2R^2}+\frac{{\Lambda}aR^n}{6}(1+\frac{n\Delta{R}}{2R}+\frac{n(n-1)\Delta{R}^2}{6R^2}) \nonumber\\
           &+m\Lambda(R^{n-1}(1+\frac{(n-1)\Delta{R}}{2R})(\frac{2a}{3}+\frac{1}{6}) \nonumber\\
           &-\frac{(a+1)R^{n-1}}{6}(1+\frac{n\Delta{R}}{2R})+\frac{R^{n-2}(a+1)\Delta{R}}{12})-\frac{m\Delta{R}}{2R^2}]
\end{eqnarray}

The observer at B will experience that coordinate time interval in proper time to be measured by the observer at B between transmission and reception of the signal is given by,

\begin{eqnarray}
&{\Delta\tau_1}=(1-\frac{2m}{R}-\frac{a\Lambda{R}^n}{3})^{1/2}\Delta{t} \nonumber\\ 
            &\simeq2L_{BC}[1+\frac{{\Lambda}aR^n}{6}(\frac{n\Delta{R}}{2R}+\frac{n(n-1)\Delta{R}^2}{6R^2}) \nonumber\\ 
    & +m\Lambda(R^{n-1}(1+\frac{(n-1)\Delta{R}}{2r})(\frac{2a}{3}+\frac{1}{6})-\frac{R^{n-1}(a+1)}{6}(1+\frac{n\Delta{R}}{2R}) \nonumber \\ 
    &+\frac{R^{n-2}(2a+1)\Delta{R}}{12}-\frac{aR^{n-1}}{6}(3+\frac{n\Delta{R}}{2R}))-\frac{m\Delta{R}}{2R^2}]
\end{eqnarray}

In deriving the above equations, the higher order terms in $\Lambda$ and $m^2\Lambda$, $m^3$, $m^2\frac{\Delta{R}^2}{R^2}$and higher order m terms have been neglected.  

\section{Discussion and Conclusion}

Dark energy has a significantly different kind of influence on gravitational time advancement than that of pure Schwarzschild geometry. The time advancement effect is entirely due to pure Schwarzschild geometry, while dark energy leads to only time delay effect which means gravitational time advancement effect will be reduced in presence of dark energy. When $\Lambda r_{A}^{2} > 2\mu/r_{B}$, there no time advancement at all. So in principle the time advancement effect should be able to identify dark matter clearly. 

In contrast the conformal theory description of flat rotation curve suggests large time advancement effect. The fitting of galactic rotation curves suggest $\Lambda/3 = -\left(5.42 \times 10^{-42} \frac{M}{M_{\odot}} + 3.06 \times 10^{-30}\right)$ $cm^{-1}$ [60]. Therefore, in our galaxy, dark matter potential should start dominating over the Schwarzschild part at distances larger than about 30 kpc. Hence at distances beyond $\sim 30$ kpc time advancement effect will be quite large. The experimental realization to examine gravitational time advancement effect at such distances is a challenging issue.

Here it is worthwhile to mention that the gravitational time advancement effect has not been experimentally verified yet but it should not be very difficult to test the effect. This is because the magnitude of time advancement effect is reasonably large. In fact gravitational time advancement is a much stronger effect than gravitational time delay when large distances are involved. However, time delay has the advantage of probing stronger gravity. In the solar system tests of gravitation, time delay measurements are mainly relied on passage of radiation grazing the sun thereby solar gravitational potential at the surface of the sun comes into play. In such a situation the time delay is about 240 $\mu sec$ whereas the total special relativistic travel time between the earth and the sun is about 1000 seconds which means the gravitational time delay is about $2 \times 10^{-7}$ part of the total travel time. For testing gravitational time advancement from the earth or its surroundings, on the other hand, solar gravitational potential at the position of earth shall be applicable and when light propagates from the earth to say Pluto and back, the time advancement will be about 1 msec over the total propagation time of 50000 seconds i.e. here the time advancement is about $0.2 \times 10^{-7}$ part of the total travel time which is just one order smaller than time delay caused by the sun and hence is detectable. Note that the above estimates need to be corrected taking into account the variations in round-trip travel time due to the orbital motion of the target relative to the Earth by using radar-ranging or any other similar kind of data. Since gravity can not be switched off, one does not have access to a special relativistic propagation of photon against which the time delay to be measured. Therefore, variations of time delay is measured as a function of distance to verify the radial profile of Eq.(3). Similar check can be made for the time advancement also.

The future missions such as the Beyond Einstein Advanced Coherent Optical Network (BEACON) [61] or the GRACE Follow-On (GRACE-FO) mission [62] will probe the gravitational field of the Earth with unprecedented accuracy. The BEACON mission will employ four small spacecraft equipped with laser transceivers and the spaceraft will be placed in circular Earth orbit at a radius  of $80,000$ km. All the six distances between the spacecraft will be measured to high accuracy ($\sim 0.1 \; nm$) out of which one diagonal laser trajectory will be very close to the Earth thereby pick the gravitational time delay effect. If the distance between the spacecraft and the Earth is also measured by an Earth bound observer and compared with distances measured by the spacecraft, the time advancement effect may be revealed from the measurements. The GRACE-FO, which is scheduled for launch in 2017, will be equipped with a laser ranging interferometer and is expected to provide range with an accuracy of $1$ nm and with such level of accuracy general relativistic effects may become significant [63]. It is, threrefore, important to examine whether the effect of time advancement can have any significant effect on observables of GRACE-FO. 

To probe dark matter through it's influence on gravitational time advancement properly, one requires to observe time advancement (delay) effect at distance $\sim 30$ kpc or beyond. For probing dark energy observations are to be made at even higher distances. This is currently not feasible. At present observations can be made only from the earth or from its neighborhood via a satellite/space station. So strategies to be developed  for observing time advancement/delay effect at other distances, may be some indirect means. This would be very challenging task. 

For small distance travel, the time advancement effect is a second order effect, unlike the long distance travel where time advancement occurs due to first order effect. However, since the time advancement effect is cumulative in nature, if a light beam is allowed to travel say from the earth surface radially upwards to a nearby point large number of times it (the light beam) should acquire time advancement of reasonable magnitude when observed from the earth surface which should be measurable.   

In summary, we investigate the influence of dark matter/energy on gravitational time advancement. We obtain analytical expression for time advancement to first order in $M$ and $\Lambda$ where $\Lambda$ is the parameter describing the strength of dark matter/energy. From our results it is found that dark energy leads to gravitational time delay only whereas pure Schwarzschild metric gives both time delay and time advancement (negative effective time delay) depending on the position of the observer. 

The present finding suggests that in principle the measurements of gravitational time advancement at large distances can verify the dark matter and a few dark energy models or put upper limit on the dark matter/energy parameter.

\end{document}